\documentstyle[12pt,float,psfig]{article}
\restylefloat{figure}
\restylefloat{table}

\advance \topmargin by -\headheight
\advance \topmargin by -\headsep
\evensidemargin \oddsidemargin
\title{Modeling of Oscillatory Driven Surface Evolution}

\author{M. Khenner
\vspace{0.5cm}\\
Department of Mathematics \\
State University of New York at Buffalo \\
Buffalo, NY 14260\\
email: mkhenner@nsm.buffalo.edu}

\newcommand{\rf}[1]{(\ref{#1})}
\newcommand{\beq}[1]{ \begin{equation}\label{#1} }
\newcommand{\eeq}{\end{equation} }

\begin{document}
\pagestyle{plain}
\maketitle

\begin{abstract}

\noindent


A continuum (Mullins-type) model is formulated for the isotropic evolution of a solid surface 
on which the mass transport occurs by oscillatory surface diffusion.
The time-space oscillations of diffusivity are assumed to be induced by external source(s);
a number of possible sources, different in their physical nature has been suggested 
in \cite{P-LH}. The particular driving mechanism assumed here is low-energy, pulsed electron or
laser beam(s) incident on a surface. 
The simplified version of the model is applied
to study the relaxation of a periodic surface corrugation. 
Several extensions of the model are proposed. 



\end{abstract}

Considered in this Letter is a model 1+1 case corresponding to a two-dimensional crystal with a 
one-dimensional surface.
The evolution of the crystal surface given by  $z=h(x,t)$ is described by the
following phenomenological partial differential equation \cite{MULLINS57,TCH94}:
\begin{equation}
\frac{h_t}{(1+h_x^2)^{1/2}} = V = - \Omega j_s,
\label{1.1}
\end{equation}
where $V$ is the normal velocity of the curve representing the surface,
$\Omega = const.$ the atomic volume, $j$ the surface diffusion flux, and
$s$ the arc length along the curve. 
Subscripts denote differentiation in all equations in the Letter.

The surface flux $j$ is given by 
\begin{equation}
j = -\frac{D \nu}{kT} \mu_s,
\label{1.2}
\end{equation} 
where $D$ is the surface diffusivity
of adatoms, $\nu=const.$ the surface density of adatoms, $k$ the Boltzmann constant,
$T$ the temperature, and 
\begin{equation}
\mu = \Omega \gamma \kappa
\label{1.3}
\end{equation} 
the surface chemical potential.
In \rf{1.3}, $\gamma$ is the isotropic surface energy density and
\begin{equation}
\kappa = \frac{-h_{xx}}{(1+h_x^2)^{3/2}}
\label{1.3a}
\end{equation}
the surface mean curvature. Note that more complicated expressions for $\mu$ must be used 
in the cases of weakly and strongly anisotropic surface energy density \cite{GoDaNe98}.

\section{Oscillatory driving of surface diffusion by pulsed electron or laser beams}

\label{Sec2}

It was recently pointed out in \cite{P-LH} that ``oscillatory driving appears
as an alternative route for pattern formation, with two basic advantages: First, patterning and growth are
separated, so that morphology is not a function of the growth process. Second, it offers better control of 
the structure. An \textit{in situ} and real-time control of the pattern becomes possible, opening
a wide range of new applications." 
On the side of fundamental physics, oscillatory driving of solid surfaces falls in the scope 
of intensively studied nonlinear phenomena such as the inverted pendulum problem,
the Faraday instability of a fluid surface, and pattern formation in granular media 
(see references in \cite{P-LH}).

Work \cite{P-LH} employs a mesoscopic step-flow model for the surface to calculate the induced mass 
flux along vicinal
surfaces, which accounts for surface dynamics and stability. Slope selection, surface metastability,
and driving frequency-dependent surface stability are found.  
No particular \textit{physical} driving mechanism was assumed in \cite{P-LH}, and  only
the surface diffusivity oscillating in time but uniform in space was considered.

Let pulsed, low-energy photon or, more often, electron beam is incident on a solid surface (`low-energy' means
that the incident energy is insufficient to cause the removal of atoms from the surface, through
elastic collision). 
The collision energy generates
an excited surface state, often free carriers (electrons and holes), which recombine with the local release
of heat.
Two types of enhanced surface diffusion are usually 
distinguished (\cite{IS}, pp. 250-254). 
One mechanism is `local heating'
or 'hot-spot'; the key step here is the conversion of the recombination energy into local vibration energy 
of atoms. Some of the recombination energy and some  thermal energy are used to achieve a passage 
over the
potential barrier to motion. Another mechanism is 
`local excitation' and it makes use of an energy surface corresponding to an excited electronic 
state.
In reality, at least when laser beams (photons) are incident on a semiconductor surface 
both mechanisms are accompanied by 
`direct' enhancement,
that is by conversion of incident energy into lattice vibrational energy through collisional
momentum transfer. 

In most cases, one recombination event
can produce at most one diffusion jump. This means that rather large number of recombinations is needed
to cause significant  matter transport. 
For the atomic displacements through collisional momentum transfer,
the energy of electrons should be higher than about 100 keV; for displacement through electronic excitation,
the electron energy could be as small as 5 eV (9 eV for GaAs; see Table 2.1, p. 60 in \cite{IS}).
Publications \cite{DS}-\cite{NIKK} discuss surface effects 
caused by electron and laser beams (including the diffusion).

On a macroscopic level of description used here, pulsed radiation gives rise to 
a quasistationary state in which the temperature at the targeted spot fluctuates about the
mean value $T_0$ with a frequency equal to the source pulse repetition frequency $\omega$,
and an amplitude $\tilde T(x)$, where $x$ is coordinate across the spot \cite{Yakunkin}. 
The value of $T_0$ and the functional form of $\tilde T(x)$ are 
determined by the impulsive power density $I$ and the mean power density $\bar I = \omega t_i I$
of the radiation ($t_i$ is the pulse duration), the absorbivity of the surface at the 
radiation wavelength, and the
thermophysical characteristics of the material. Thus the surface diffusion can be, at least in
principle, controlled
through temperature oscillations, and I consider this mechanism in the following 
sections.

\section{Derivation of the surface evolution equation}

\label{Sec3}

Let the surface diffusivity is given by the familiar Arrhenius expression 
\begin{equation}
D(T) = D_0 \exp{\left(\frac{-E_d}{kT}\right)},
\label{1.15}
\end{equation}
where $D_0=const.$ and $E_d=const.$ are the pre-factor and activation energy, respectively. 

Substituting the equations \rf{1.2} - \rf{1.15} into the equation  
\rf{1.1}, and using $\partial/\partial s = (1+h_x^2)^{-1/2}\partial/\partial x$ I obtain the following 4th-order surface evolution equation:
\begin{equation}
h_t = B\frac{\partial}{\partial x}\left(\frac{\exp{\left(\frac{-E_d}{kT}\right)}}{kT}
(1+h_x^2)^{-1/2} \left[\frac{-h_{xx}}{(1+h_x^2)^{3/2}}\right]_x\right),
\label{3.1}
\end{equation}
where $B = \Omega^2 \nu D_0\gamma$.

The equation \rf{3.1} admits trivial solution (base state)
\begin{equation}
h = h_0 = const.,\; T = T_0 = const. > 0,
\label{3.2}
\end{equation}
corresponding to the unperturbed, horizontal surface at elevated mean temperature. 
Without loss of generality 
$h_0 = 0$ can be chosen. Consider small perturbations of this
state:
\begin{equation}
h = \tilde h(x,t),\; T = T_0 + \tilde T(x,t),\quad \mbox {where}\; |\tilde h(x,t)|,\; 
|\hat T(x,t)| = \frac{\tilde T(x,t)}{T_0} \ll 1.  
\label{3.3}
\end{equation}
One can think of a number of cases for the perturbation 
of the mean temperature field, each of which may result in very different \textit{nonlinear} 
phenomena at the surface.
For instance: several, synchronized radiation sources are arranged in a periodic 
array along the surface. Then, the spatial shape 
of the temperature perturbation can roughly be taken periodic with the period 
equal to the separation between the beams. Or, each source has its own pulse 
repetition frequency; then, the resulting situation amounts to a multi-frequency driving
of a surface (also see Section \ref{Sec66}).

Expanding $\exp{\left(\frac{-E_d}{kT}\right)}/(kT)$ in \rf{3.1} in 
powers of $\hat T(x,t)$ and keeping only the 
zeroth-order and linear term, I find:
\begin{equation}
h_t = B a_2 \hat T_x 
(1+h_x^2)^{-1/2} \left[\frac{-h_{xx}}{(1+h_x^2)^{3/2}}\right]_x + B (a_1+a_2 \hat T)
\left\{(1+h_x^2)^{-1/2} \left[\frac{-h_{xx}}{(1+h_x^2)^{3/2}}\right]_x\right\}_x,
\label{3.4}
\end{equation}
where 
\begin{equation}
a_1 = \frac{\exp{\left(\frac{-E_d}{kT_0}\right)}}{kT_0},\quad a_2 = a_1\left(\frac{E_d}{kT_0}-1\right).
\label{3.41}
\end{equation}

Next, the equation \rf{3.4} is expanded in powers of $h_x$ up to second order; 
considered are only such perturbations of the temperature field that 
$|\hat T_x| < |h_x| \ll 1$ (so that terms 
proportional to $\hat T h_x, \hat T_x h_x$ and higher order are ignored).
This procedure yields the following equation:
$$
h_t = B\left[-(a_1 + a_2 \hat T) h_{xxxx} -
a_2 \hat T_x h_{xxx} + 3 (a_1+ a_2 \hat T) h_{xx}^3 +\right.
$$
\begin{equation}
\left.10a_1h_{xx}h_{xxx}h_x +
a_1\left(2h_{xxxx}-27h_{xx}^3\right)h_x^2 \right] + O(h_x^2).
\label{3.7}
\end{equation}
The equation \rf{3.7} is the starting point for studies of effects of temperature 
variations on the surface diffusion-produced morphologies of solid surfaces. 
More terms of the expansion must be kept in this equation if the condition of small slope of the  temperature perturbation is relaxed. The key simplifying assumption in the derivation is 
the independence of the surface energy density on temperature, and its isotropy.

\section{Case of spatially uniform temperature field}

\label{Sec4}

Let the temperature perturbation is the time-periodic function, that is
$\hat T(x,t) = \hat T(t) = T\cos{\omega t}, 0 < T \ll 1.$ The equation \rf{3.7} becomes
\begin{equation}
h_t = -B\left[\left(a_1+ a_2T\cos{\omega t}\right)\left(h_{xxxx}-
3h_{xx}^3\right)-
10a_1h_{xx}h_{xxx}h_x -
a_1\left(2h_{xxxx}-27h_{xx}^3\right)h_x^2 \right].
\label{3.10}
\end{equation}
Notice that \rf{3.10}, as well as the more general equation \rf{3.7} 
is nonlinear even in the zeroth order of small-slope expansion.
Simple linear approximation is now considered
(all spatial derivatives $\ll 1$). 

In the linear approximation, \rf{3.10} reduces to
\begin{equation}
h_t = -\hat B\left(1+ A\cos{t}\right)h_{xxxx},\quad \hat B = \frac{a_1B}{\omega L^4} > 0,
\quad A = \frac{a_2}{a_1}T.
\label{3.101}
\end{equation}
Equation \rf{3.101} is in the nondimensional form; $1/\omega$ is used for unit of time
and the diameter of the irradiated spot, $L$, for unit of length. Notice that $A$ is positive 
if $E_d>kT_0$ and negative otherwise.

Solution of the equation \rf{3.101} implies that the initial surface corrugation 
with the shape of  $\cos{\alpha x}$ (where $\alpha$ is
arbitrary wave number) relaxes to the base state $h=0$ in oscillatory fashion as
\begin{equation}
\exp{\left(-\hat B\alpha^4 t\right)}\exp{\left(-\hat BA\alpha^4\sin{t}\right)},
\label{3.11}
\end{equation} 
while preserving the initial shape.
Such relaxation mode should be contrasted to monotonous exponential decay as
\begin{equation}
\exp{\left(-\hat B\alpha^4 t\right)}
\label{3.12}
\end{equation} 
in the absence of oscillatory driving, that is at constant $T=T_0$ (here, for comparison with the
case of oscillatory driving, the time and length units are chosen same).
At typical values $L=100$ nm, $\Omega=2\times10^{-23}$ cm$^3$,
$\gamma=10^3$ erg/cm$^2$, $\nu=10^{15}$ cm$^{-2}$,  $\omega=5\times 10^4$ s$^{-1}$, 
$D_0=10^{12}$ cm$^2$/s, $E_d=0.5$ eV, $T_0=300$ K, $T=10^{-2}$, one obtains $\hat B = 4\times 10^4, A = 0.2$. Thus the corrugation relaxes fast unless it has long wave length; with the diameter
of the irradiated spot or pulse repetition frequency increased, shorter wave length perturbations live longer.

\section{Discussion and conclusions}

\label{Sec66}

This Letter suggests the simple, 1+1 continuum model of the driven evolution of the morphology 
of a crystal surface irradiated by the pulsed, low-energy electron or laser beam(s). 
The solution to the evolution equation has been obtained analytically in the particular case of spatially uniform field of temperature oscillation on a surface, and the small-slope, linear
approximation.

A number of future directions for the research can be identified.

\begin{enumerate}

\item Study of the evolution of the surface morphology within the context of 
nonlinear, small-slope
equations \rf{3.10} and \rf{3.7}, as well as of the equation \rf{3.4} (which is applicable
for large slopes); the latter two equations require an assumption on the
form of spatial dependence of the temperature perturbation.
These equations are unsolvable analytically if no additional simplifying assumptions are made.
Therefore, one has to resort to numerical methods.
It is expected that the nonlinearity will support the oscillations and slow down the rate 
of relaxation
of the corrugation with the emergence of quasi-stationary pattern. 
Its structure and dependence on driving frequency and amplitude is 
of prime interest.

\item Derivation, analysis and solution of the surface evolution equation 
under assumption that temperature oscillations exert influence on the anisotropic
surface energy density only.
Note that during crystal growth with anisotropic surface energy (no oscillatory driving) 
and in the longwave limit a \textit{stationary} solutions in the form of a periodic hill and valley 
structures exist \cite{SGDNV}; the oscillatory driving is expected to produce a different
quasistationary pattern.
The model that is most relevant to practice would include the temperature dependence of both the
diffusivity and the surface energy density.

\item The drawback of the model presented here is the \textit{ad hoc} inclusion of the 
temperature perturbation. To make the model more realistic, one should
formulate and solve the problem for the thermal diffusion on the 
irradiated surface \cite{Yakunkin,Bennett}. 
The solution of that problem, $T(x,t)$
must then be used for the temperature perturbation.

\item Modeling of the crystal growth. 
Assume for a moment that the oscillatory driving of a crystal surface through the
pulsed radiation mechanism discussed in Section \ref{Sec2} is absent.
Then the impinging oscillatory flux of material particles on a surface 
will not result in new instabilities but may change the thresholds and
resulting patterns arising from known instabilities during crystal growth with non-oscillatory 
flux \cite{P-LH}. However, the two oscillatory mechanisms acting simultaneously may create
completely new instabilities. This situation bears resemblance to two-frequency
driving of a fluid surface, which is an actively studied subject \cite{ZV} (of course, physical 
media, conditions, mechanisms and governing equations are completely different).

\item Extension of the above models to second spatial dimension. 

\end{enumerate}

I'm currently working on the first two directions.
The long-term impact of the mathematical modeling of externally driven 
surfaces to the technologically important field of studies of surface nanoscale 
patterns and their applications is expected.


\end{document}